\documentclass[twocolumn,amsmath,amssymb]{snp}
\usepackage{graphicx}
\usepackage{dcolumn}
\usepackage{bm}
\newcommand{\nn}{\nonumber}
\newcommand{\be}{\begin{equation}}
\newcommand{\ee}{\end{equation}}
\newcommand{\bea}{\begin{eqnarray}}
\newcommand{\eea}{\end{eqnarray}}

\newcommand{\om}{\omega}

\newcommand{\vp}{\vec p}

\begin{document}

\title{{\Large  Fluid property of QGP in presence of magnetic field}}

\author{\large Souvik Paul$^{1,*}$, Ankita Mishra$^2$, Jayanta Dey$^3$, Sarthak Satapathy$^3$,
Sabyasachi Ghosh$^3$}
\email{sp17ms070@iiserkol.ac.in}
\affiliation{$^1$Department of Physical Sciences,Indian Institute of Science Education 
and Research Kolkata, Mohanpur, West Bengal 741246, India}
\affiliation{$^2$Department of Mechanical Engineering, Guru Ghasidas University, Bilaspur 495009, India}
\affiliation{$^3$Indian Institute of Technology Bhilai, GEC Campus, Sejbahar, Raipur 492015, 
Chhattisgarh, India}
\maketitle

Present work has gone through two parts of investigations. First, it attempts
to map interaction of quantum chromo dynamics (QCD) in presence of external magnetic field,
described from lattice QCD (LQCD) calculation~\cite{Bali2}. Then it explore the fluid
property of quark gluon plasma (QGP) in presence of external magnetic field by
using the mapped interaction, which depends on temperature $T$ and magnetic field $B$.
\begin{figure}
 \centering
 \includegraphics[scale=0.5]{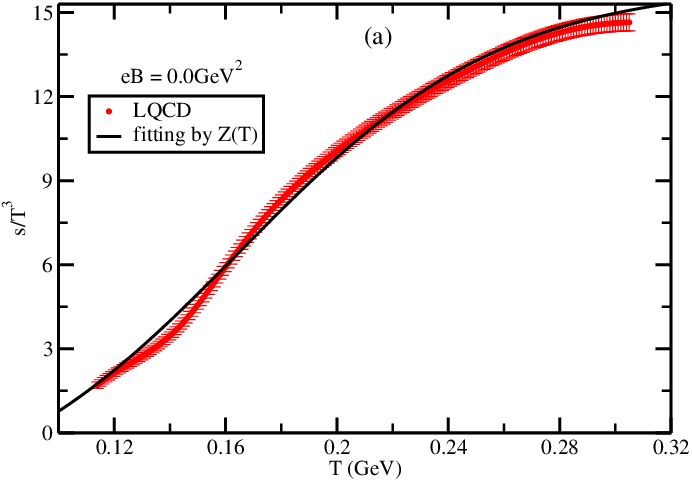} 
 \includegraphics[scale=0.24]{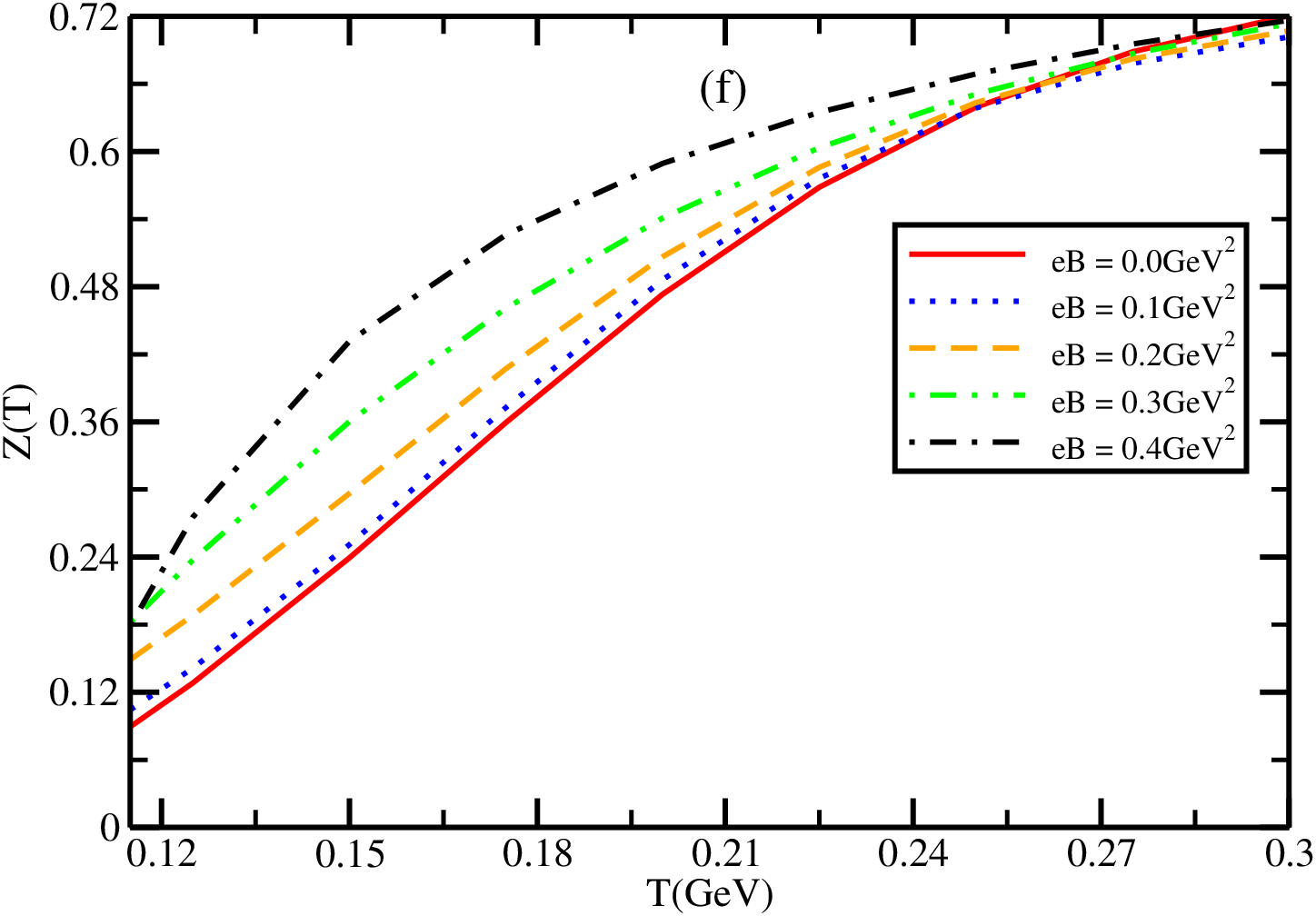} 
 \caption{Fitting curves (black solid line) and LQCD data points (red circles) for normalized 
 entropy density $s/T^3$ by parameterizing $Z(T)$ at different 
 magnetic field strengths - (a) $eB=0$, (b) $0.1$, (c) $0.2$, (d) $0.3$ (e) $0.4$ GeV$^2$.
 Corresponding $Z(T)$ curves are plotted in (b).}
 \label{ZvsT}
 \end{figure}

Let us start with quasi-particle expression of entropy density for non-interacting QGP,
\bea
s &=& \sum_{i=u,d,s,g}g_i
\int_{0}^{\infty}\frac{d^3p}{(2\pi)^3}
\nn\\
&&\Big(\om_i+\frac{p^2}{3\om_i}\Big)f_i~,
\label{s_QGP}
\eea
where $f_i$ is Fermi-Dirac/Bose-Einstein distribution functions of fermions/bosons (quarks/gluons)
and $\om_i=\sqrt{\vp^2+m_i^2}$ with $m_{g,u,d,s}=0$, $0.005$, $0.005$, $0.100$ GeV. 
The lattice Quantum Chromo Dynamics (LQCD) 
simulation~\cite{Bali2} obtained always lower values of $s$ with respect to Eq.~(\ref{s_QGP}),
which is approximately $s\approx 20.8~T^3$, commonly called Stephan-Boltzmann (SB) limit.
In this context, we have followed the prescriptions of Chandra and 
Ravisankar~\cite{Chandra_PRC07}, where QCD interaction of quark-hadron phase transition has 
be mapped via fugacity parameter $Z$. It has nothing to link with quark or gluon
chemical potentials, which are absolutely kept zero. The $Z$ enters in $f_i$ as 
\be
f_i=\frac{1}{Z^{-1}{\rm exp}\Big(\beta \om_i\Big)-a_i}~,
\ee
where $a_i=\pm 1$ for fermions/bosons (quarks/gluons) and $Z=1$, $Z<1$ represent 
non-interacting and interacting QGP system respectively.
The LQCD data of $s(T)$ for different $eB$'s are re-drawn red circles in Fig.~\ref{ZvsT}(a).
Then, they are fitted (solid black lines) by assuming a temperature dependent fugacity $Z(T)$, 
which are plotted in Fig.~\ref{ZvsT}(b).
So by reducing fugacity with reduction of
temperature or/and magnetic field, one can transform a non-interacting to
interacting QGP system description and we found a gross $Z(T, B)$ function, which
can nicely map the $T$ and $B$ dependent QCD interaction, provided by LQCD~\cite{Bali2}. 

After building the quasi-particle model of interacting QGP in presence of magnetic field,
we have estimated normal ($\eta_2$) and Hall ($\eta_4$) shear viscosity components in 
presence of magnetic field from the standard relations~\cite{JD_eta}
\bea
\eta_{2,4} &=& \sum_{i=u,d,s,g}\frac{g_i\beta}{15}\int\frac{d^3\vp}{(2\pi)^3}\frac{\vp^4}{\om_i^2}
f_i(1-a_i f_i)
\nn\\
&&~~~~~~~~~~~~~~~~~~~~~\frac{\tau_c(\tau_c/\tau_B)^{0,1}}{1+(\frac{\tau_c}{\tau_B})^2}~,
\label{eta_24}
\eea
where $\tau_B=\omega/(eB)$ is appeared as a new time scale due to magnetic field
along with the relaxation time $\tau_c$, already existed at $B=0$. Since gluon $g$
is charge neutral ($e=0\Rightarrow \tau_B=\infty$), it will follow without field
($B=0\Rightarrow\tau_B=\infty$) expression of $\eta$, which can be realized as 
$\eta=\eta_2(B=0)$.
\begin{figure}
 \centering
 \includegraphics[scale=0.23]{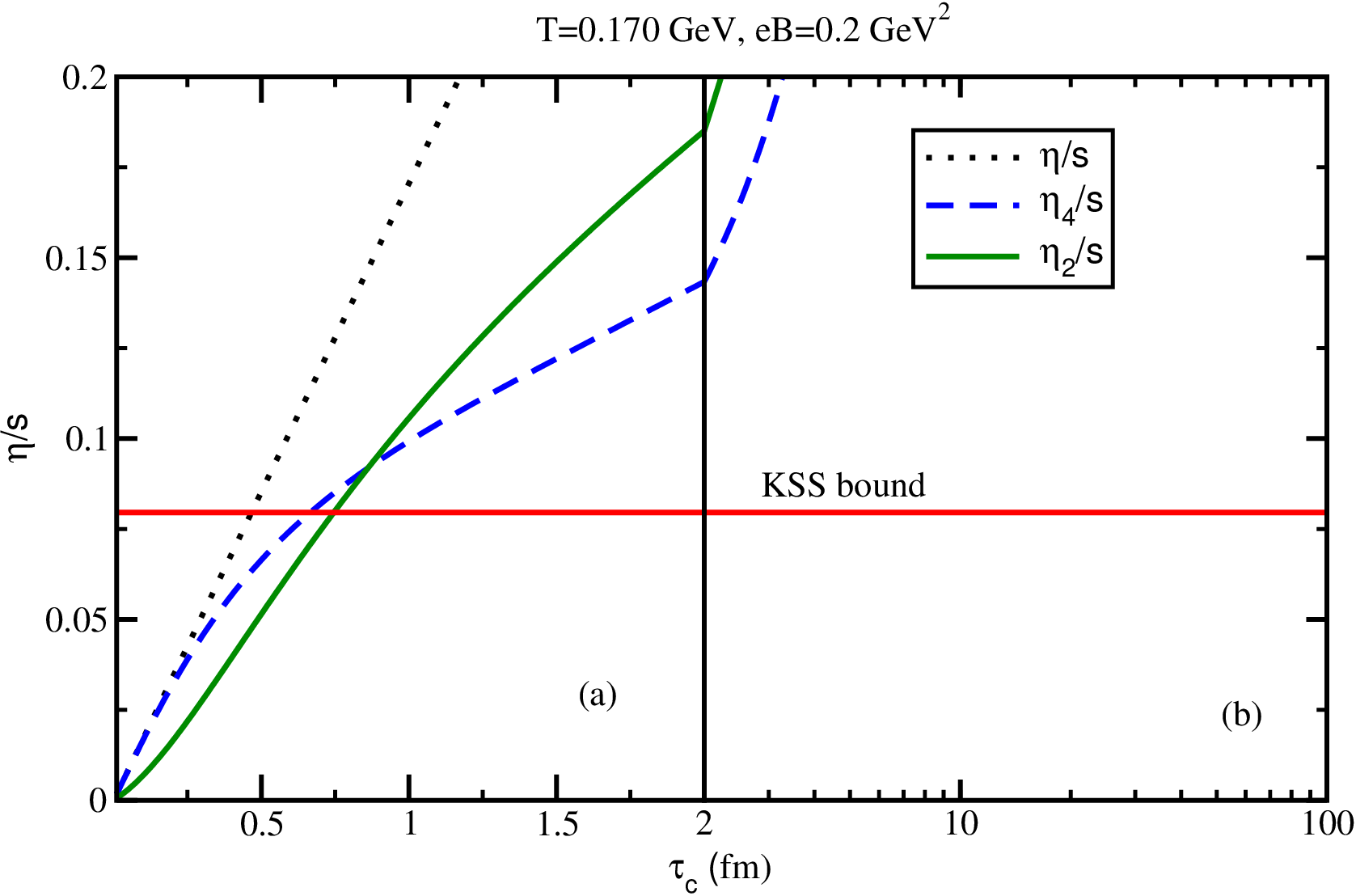}
 \caption{Collisional relaxation time ($\tau_c$) dependence of shear viscosity to entropy density ratio.}
 \label{es_tc}
 \end{figure}

Shear viscosity to entropy density ratio measures the fluid nature of the medium. From experimental
side, this quantity should be very close to KSS bound $1/(4\pi)$~\cite{KSS}, which is
drawn by red solid (horizontal) line in Fig.~(\ref{es_tc}). With the help of 
Eqs.~(\ref{eta_24}), (\ref{s_QGP})
we have estimated $\eta_2/s$ (solid line), $\eta_4/s$ (dash line), $\eta/s=\eta_2(B=0)/s$ (dotted line),
which are plotted against $\tau_c$-axis in Fig.~(\ref{es_tc}). 
Analyzing Eq.~(\ref{eta_24}), one can recognize $\frac{\eta}{s}\propto\tau_c$ for all quarks 
and gluons at $B=0$, while at finite $B$, $u, d, s$ quarks will acquire anisotropic viscosity components
$\frac{\eta_{2,4}}{s}\propto\frac{\tau_c(\tau_c/\tau_B)^{0,1}}{1+(\tau_c/\tau_B)^2}$ 
but gluon component remain isotropic. Hence, the proportional curve of $\eta/s$ is slightly 
bended due to finite $B$ as represented by $\eta_{2,4}/s$ curves. The bending is appeared
to be mild as gluon's $\propto\tau_c$ contribution is added with quark's
$\propto\frac{\tau_c}{1+(\tau_c/\tau_B)^2}$ contribution. Excluding gluon component,
one can see that $\eta_{2}/s$ first increases and then decreases with $\tau_c$, where
their peak can be seen around $\tau_c\approx\tau_B$. therefore, this $\eta_2/s$ can
cross the KSS line at two times in $\tau_c$-axis. This fact is well explored in Ref.~\cite{JD_eta}.
As normal component $\eta_2$ decreases and $s$ increases with $B$, so $\eta/s$ decreases
more in presence of magnetic field. Hence, Present investigation recognize two sources - magnetic field
and interaction, which can push the QGP system towards its nearly perfect fluid nature, quantified 
by KSS line. A detail investigation on it can be seen in Ref.~\cite{QGP_B}.

\end{document}